# High throughput screening for LC3/GABARAP binders utilizing the fluorescence polarization assay

**Martin P. Schwalm** [1,2]†, **Johannes Dopfer**[1,2]†, **Stefan Knapp** [1,2], **Vladimir V. Rogov**[1,2]*

[1]Institute of Pharmaceutical Chemistry, Goethe University Frankfurt,

Max-von-Laue-Str.9, 60438 Frankfurt, Germany

[2]Structural Genomics Consortium, BMLS, Goethe University Frankfurt,

Max-von-Laue-Str. 15, 60438 Frankfurt, Germany

*Correspondence: rogov@pharmchem.uni-frankfurt.de

† contributed equally

Abstract

The characterization of interactions between autophagy modifiers (Atg8-family proteins) and their natural ligands (peptides and proteins) or small molecules is important for a detailed understanding of selective autophagy mechanisms and for design of potential Atg8 inhibitors that affect the autophagy processes in cells. The fluorescence polarization (FP) assay is a rapid, cost-effective and robust method which provides affinity and selectivity information of small molecules and peptide ligands targeting human Atg8 proteins.

This chapter introduces the basic principles of FP assays are introduced. In addition, a case study for FP small molecule and peptide interactions with human Atg8 proteins (LC3/GABARAPs) is described.

Finally, data analysis and quality control of FP studies are discussed for the proper calculation of $K_i$ values of the measured compounds.

Key words

fluorescence polarization, $K_i$ determination, LC3, GABARAP, LIR motif, High Throughput Screening

1 Introduction

Autophagy is a conserved cellular pathway that aims to degrade and recycle variety of cellular components (protein aggregates, cellular organelles, invaded bacteria, etc.) and thus maintain cellular homeostasis (reviewed in [1, 2]). Autophagy is manifested by the double-membrane vesicle – autophagosome, which initiation, maturation and fusion with lysosomes/vacuoles are controlled by ~40 autophagy-related (Atg) proteins [3, 4]. Among these, ubiquitin-like Atg8 family proteins, including human LC3A, LC3B, LC3C, GABARAP, GABARAPL1, and GABARAPL2 proteins (LC3/GABARAP hereafter), play an essential role. They are the key regulators of autophagosome biogenesis and also mediate plethora of protein-protein interactions – with core Atg's, with autophagy receptors and with autophagy cargo [5-7]. Recent investigations have revealed that LC3/GABARAPs interact with their partner proteins via relatively short LC3-interation region (LIR) motifs, typically located in unstructured regions of autophagy receptors and other LC3/GABARAP-interacting proteins [8, 9]. LC3/GABARAPs provide a distinct interaction surface, called LIR docking site (LDS) which served for the specific recognition and binding of most functional LIR motifs. The LDS consists of two hydrophobic pockets (HP1 and HP2), formed by residues of a central β-strand and the adjacent α-helices within LC3/GABARAPs. The LIR motif from the most investigated autophagy receptor p62/SQSTM protein (p62 thereafter) has low micromolar affinity to the LC3/GABARAP proteins and it displays no significant selectivity to any of the human Atg8 orthologues [10]. Another interacting surface (UDS, UIM docking site, [11]) is located at the opposite side of LC3/GABARAP molecules but the UDS-mediated interactions with endogenous interaction partners have not been characterised structurally and functional yet.

The interactions between LC3/GABARAP and their interaction partners are still in the focus of modern research in autophagy. Variety of complex biophysical methods have been implemented to investigate the affinity and selectivity of these interactions (reviewed in [9]). However, each method has its own experimental limitations and taken alone it is not able to provide complete and reliable information on interactions involving LC3/GABARAP with their binding partners. Reliable methods are especially important to verify recently developed small molecules binding to the LC3/GABARAPs that may serve for the development of autophagy-related protein degraders as to dated targeted protein degradation (TDP) approaches are mainly based on small molecules that recruited E3 ligases [12]. Due to the key role of LC3/GABARAP proteins in the recruitment of cargo for the lysosomal degradation, a similar mechanism has been proposed exploiting LC3/GABARAP proteins for TPD [13]. Despite the interest of LC3/GABARAP-interacting small molecules either as autophagy modulators/inhibitors or ligands for the development of chimeric molecules for TPD, screening campaigns have only yielded few low affinity binder [14]. Hence, more screening campaigns are required for the identification of LC3/GABARAP-binders together with a high throughput method for affinity and selectivity determination. This protocol describes the use of tracer-based fluorescence techniques for the screening and characterization of compounds for the identification of LC3/GABARAP inhibitors.

Fluorescence polarization is based on the relation between the hydrodynamic radius of a molecule in solution and the Brownian motion of it. Light emitted by a fluorophore (e.g. Cy5) after excitation with polarized light, will be depolarized to a certain extend. Since the exited state of fluorophores possesses a limited lifetime, the rotational speed of this fluorophore influences the level of depolarization (Fig. 1a). Therefore, increasing the mass of the fluorophore complex will decrease the rotational speed together with the amount of depolarization and it is therefore measureable (Fig. 1b). This principle can be utilized by fluorescent labelling of a peptide (here, we used 17-meric peptide contained prototypical p62-LIR motif) which is known to bind to the protein of interest (in our case all 6 human LC3/GABARAP proteins) in low micromolar $K_D$ region as shown in Fig. 1c. The small Cy5-labelled p62-LIR peptide in its unbound state has a much faster rotational speed (tumbling) as compared to the LC3/GABARAP-bound state where its mass is drastically increased. The mass increase upon binding slows down the tumbling of the attached fluorophore and allows measurements of the peptide binding to LC3/GABARAP proteins by tracking the amount of depolarized light emission. To implement this principle for ligand identification, displacement experiments reveal inhibitory constant values ($K_i$) for each tested ligand, allowing affinity determination of ligands binding to the LDS of LC3/GABARAP (as the p62-LIR interacts to LC3/GABARAPs exclusively by LDS). Subsequent comparison of the $K_i$ values for all 6 human LC3/GABARAPs will provide the selectivity profile of the measured ligand. Despite it robustness, the fluorescence polarization assay has some limitations, including a problems of very high and very low ligand affinity, ligand-induced polarization quenching and interference, non—specific ligand binding, etc. Therefore, the FP method should be implemented carefully with a number of positive and negative controls.

This protocol describes, how to purify all 6 human Atg8 homologues from *E. coli* cells to subsequently use these proteins for the establishment of fluorescence polarization binding assays for high throughput ligand screening and binder characterization.

## 2 Materials

All compounds were dissolved in 100 % DMSO to a final volume of 10 mM. Chemicals and consumables are used as indicated the first time mentioned in the text and described with supplier and catalogue number. Expression plasmids of LC3A, LC3B, LC3C, GABARAP, GABARAPL1 and GABARAPL2 are available upon request. Peptides were synthesized by GeneCust while compounds were ordered from Sellechem, Tocris, MedChemExpress or Sigma Aldrich.

# 3 Methods

## 3.1. Protein expression and purification

Since fluorescence polarization is a biophysical method for *in vitro* characterization of protein:ligand interactions, purified protein must first be obtained. For this, plasmids containing an open reading frame of individual human Atg8 homologues were used. These plasmids contain a T7 promoter for protein expression in *E. coli* Rosetta BL21 (DE3) cells and a TEV-protease cleavable $His_6$-tag. Protein expression and purification was carried out as described in the following:

1. Transforming plasmids: Thaw 50 µl of *E. coli* Rosetta (DE3) cells (Sigma-Aldrich, Cat# 71400-3) on ice and add 1-10 ng plasmid DNA. Incubate the cells for 20 min on ice, followed by a heat shock for 30 s at 42 °C. After the heat-shock, incubate the cells on ice for additional 2-3 min before adding SOC medium (included in the Rosetta kit) and incubating the cells for 1 h at 37 °C while shaking (~300 rpm). Finally, recover the bacteria by centrifugation at 5000 g for 5 min and discard the supernatant (the supernatant should be autoclaved or sterilized chemically). Add 50 µl of fresh medium and plate the cells on LB-agar plates (Carl Roth Cat#X969.1) containing 50 µg/ml Kanamycin (Carl Roth Cat#T832.4) and 34 µg/ml Chloramphenicol (Carl Roth Cat#3886.2) followed by incubation at 37 °C overnight.

2. On the following day, use a sterile pipet tip to pick 8-10 colonies and inoculate a 100 ml culture of TB medium (Carl Roth Cat#3556.2) containing 50 µg/ml Kanamycin and 34 µg/ml Chloramphenicol. Incubate the flask at 37 °C while shaking at 180 rpm overnight.

3. Prepare 1 L of sterile TB medium for each purification using 2 L shaking flasks and add Kanamycin to a final concentration of 50 µg/ml.

4. Add 25 ml of the starter culture to each culture flask and incubate at 37 °C while shaking at 120 rpm until the cultures reach an optical density at wavelength of 600 nm ($OD_{600}$) of 0.8 using a 1 cm cuvette.

5. Cool down the culture flasks by incubation on ice or in a cold room while cooling down the shaker to 18 °C. After the culture has been cooled down, shake the cells for additional 30 min at 18 °C and 120 rpm.

6. Prepare a stock of 0.5 M of Isopropyl-ß-D-thiogalactopyranosid/IPTG (Carl Roth Cat# CN08.1) and add 1 ml to each culture flask. Continue shaking at 18 °C at 120 rpm overnight.

7. Harvest the cells by centrifugation of the cultures for 10 min at 5000 g and discard the supernatant (the supernatant should be autoclaved or sterilized chemically). Collect the cell pellet in a 50 ml reaction tube (Greiner Bio Cat#227270) which can be stored at -20 °C for future purification. For immediate purification, keep the cells on ice.

8. Resuspend the cells in lysis buffer containing 500 mM NaCl (Carl Roth Cat#0962.3), 0.5 mM Tris-(2-carboxyethyl)-phosphine hydrochloride/TCEP (Carl Roth Cat#HN95.2), 30 mM 4-(2-Hydroxyethyl)piperazine-1-ethanesulfonic acid/HEPES at pH 7.5 (Carl Roth Cat#6763.3) and 5 % glycerol (Carl Roth Cat#3783.1). Add one tablet of complete protease inhibitor (Sigma Aldrich Cat# 11836145001) together with some flakes (spatula tip) of DNase I (Sigma Aldrich Cat# 4716728001)

9. Sonicate the cell suspension on ice for 15 min with alternating 5 s on- and 10 s off-pulse. Mix the lysate by stirring followed by another round of sonication.

10. Separate the lysate from cell debris by centrifugation at 4 °C and 50000 g for 1h followed by lysate filtration using a 45 μm sterile filter (Carl Roth Cat# XA50.1).

11. Affinity chromatography: Load the filtered lysate either onto a NiNTA gravity flow column (Fisher Scientific Cat# 12856586) or use a pump system (e.g. Äkta Start, Cytiva) to load the lysate onto a NiNTA column (Cytiva Cat# 29051021). After loading the lysate, wash the column with 3 column volumes (CV) lysis buffer containing 20 mM Imidazole (Carl Roth Cat# X998.1). For gravity flow elution, add sequentially 1 CV of each lysis buffer containing 100, 200 and last 300 mM Imidazole to obtain 3 elution fractions, each 1 CV. For elution using the pump system, change the buffer to lysis buffer containing 300 mM Imidazole and monitor the A280 nm chromatogram while fractionating.

12. To evaluate the purity of recombinant proteins in each fraction, perform an SDS PAGE by loading samples (e.g. 10 µL) of the lysate before affinity chromatography, the wash and either all 3 elution fractions (gravity flow) or all elution fractions (pump system) corresponding to the peak at A280 nm (include a Protein marker for protein size estimation e.g. CozyHi™ Prestained Protein Ladder, HighQu Cat# PRL0202). Stain the gel using gel stains like e.g. ROTI®Blue quick (Carl Roth Cat#4829.1) to visualize the bands corresponding to the purified proteins.

13. To cleave the $His_6$-tag prepare 1 L of lysis buffer without imidazole (store at 4 °C). Pool fractions contained pure protein of corresponding size (try to avoid impure fractions). Add approx. 500 µg purified His-tagged TEV protease per ~50 mg of the pure proteinto the pooled protein fractions and transfer the mixture into a dialysis tube with <15 kDa cut-off pore size (Thermo Fisher Cat# 68100) which needs to be sealed carefully. Place the loaded tube in the cooled dialysis buffer and dialyse the protein while slowly stirring overnight.

14. The following day, carefully collect the liquid from the dialysis tube and load it onto a fresh NiNTA column while collecting the flow through. Wash the column with 3 CV of 30 mM Imidazole containing lysis buffer while collecting and elute with 3 CV 300 mM Imidazole containing lysis buffer while collecting the flow through.

15. To assess whether the TEV-cleavage was successful, run another SDS-PAGE gel by loading a marker, the "uncleaved" sample collected before dialysis and the three fractions from the second round of affinity chromatography. Stain the gel to visualize the bands corresponding to the purified proteins without tag by comparing the size difference between the "uncleaved" and "cleaved" fractions and isolate the collected fraction which contains most of the pure cleaved protein (pool, if possible).

16. Filter the (pooled) fraction utilizing a 0.22 µm syringe filter (Sigma Aldrich Cat#SLGL0250S) and perform a size exclusion chromatography (SEC) using SEC buffer containing 250 mM NaCl, 0.5 mM TCEP, 30 mM HEPES at pH 7.5 and 5 % glycerol. Collect all fractions showing a peak in the chromatogram at 280 nm and validate the protein size via a third SDS PAGE gel.

17. Pool all pure fractions containing the protein of expected size and concentrate the protein to approx. 2 mg/ml using a Amicon® centrifugation filter with a cut-off of 3 kDa (Sigma Aldrich Cat# UFC900308) by centrifugation at 4000 g in a swinging bucket rotor at 4 °C.

18. The protein may be stored at -80 °C (up to 1 year) after snap freezing in liquid nitrogen in aliquots of 100-200 μl.

3.2. Tracer affinity determination

Following protein purification, the affinity of the tracer has to be determined. As a tracer, the p62-LIR peptide sequence SDNSSGGDDDWTHLSSK-Cy5 (obtained from Genecust) was used. Before a reliable tracer displacement assay can be performed, the tracer $K_D$ needs to be determined for the successive calculation of inhibitory constant ($K_i$) values. Since in fluorescence polarization the tracer is the signal providing assay parameter, its concentration has to be constant, only allowing variation of the protein concentration. Therefore, a protein titration will yield polarization values, suitable for an exponential fit for the $K_D$ determination (Fig. 1d).

1. Buffer preparation: For fluorescence polarization (especially when using Cy5 as dye), the use of a detergent for the suppression of unspecific binding is often essential. Here we use 0.05 % Tween 20 (Sigma Aldrich Cat#P1379) in a buffer containing 30 mM HEPES at pH 7.5, 250 mM NaCl, 5% glycerol and 0.5 mM TCEP.

2. Determine the final tracer concentration by using different tracer concentrations in assay buffer (e.g. 30, 60, 90 nM) and evaluate which concentration yields a satisfying signal for the unbound tracer, whilst keeping the tracer concentrations as low as possible. The plate reader has to be equipped with a fluorescence polarization module for the desired dye (e.g. Cy5). Alternatively, perform the protein titration experiment with different concentrations and identify the most satisfying assay window (raw value difference between unbound and bound tracer should be >3) at the lowest tracer concentration that can still be used.

3. For protein titrations, perform a serial dilution of purified protein with assay buffer containing the tracer starting from a concentration of approx. 60 µM down to 10 nM in 10-12 steps. Transfer 10 µl of the dilutions (in triplicates) to a 384-well plate (Greiner Cat#784076) (alternatively 5 µl in a 1536 well plate (Greiner Cat# 782076)) and incubate the plate for 20 minutes at room temperature (RT).

4. Measure the plate in a microplate reader (e.g. PHERAstar FSX equipped with the FP 590-50 675-50 675-50 filter module, BMG Labtech), yielding millipolarization values (mP) consisting of values for perpendicular fluorescence intensity ($I_{perp}$) and parallel fluorescence intensity ($I_{para}$). The G-factor is an additional variable determining the experimental correction of the optics. G is either determined by the plate reader or by carrying out calibration experiments by calculating the ratio between vertically and horizontally polarized light [15]. Therefore, the mP value, reported by the plate reader is calculated using the following formula [16]:

$$mP = 1000 * \frac{I_{para} - G * (I_{perp})}{I_{para} + G * (I_{perp})} \qquad (1)$$

5. Plot the obtained mP values against the protein concentration on a linear X-axis. From this plot, the $K_D$ can be obtained by using an exponential fit. To obtain the most precise fit, the following formula can be used [17]:

$$LR = \frac{(X + L_{tot} + K_D) - \sqrt{(X + L_{tot} + K_D)^2 - 4 * X * L_{tot}}}{2} \qquad (2)$$

with:

$$L = L_{tot} - LR \qquad (3)$$

and:

$$Y = BKG + MF * L + FR * MF * LR \qquad (4)$$

Here, the variables describe the total tracer/ligand concentration (**L_tot**), while **BKG** reflects the background fluorescence of tracer without protein and **MF** describes the molar fluorescence which is determined by the division of Y units by X units. **LR** describes the concentration of the tracer/ligand bound to the protein and **L** the concentration of unbound tracer/ligand.

6. Fitting the curve with $K_D$ as a parameter yields a value for the $K_D$ describing the affinity of the tracer to the protein.

3.3. Dose-dependent peptide displacement assay for inhibitor affinity determination:

After the determination of the accurate $K_D$ value for the interaction between the target protein and tracer, displacement experiments can be performed to determine $K_i$ values for the small molecules (Fig. 2). In dose-dependent experiments, the optimal tracer concentration remains the same as described previously and the protein concentration is ideally equal to the determined $K_D$, if the $K_D$ is lower than 5 µM. If the $K_D$ is higher, use 5 µM as protein concentration to avoid excessive amounts of protein within the assay system. To obtain positive control curves, non-labelled LIR-containing peptides (e.g. p62-LIR) may be used to achieve displacement of the Cy5-labelled p62-LIR tracer (Fig. 2b).

1. Prepare an assay master mix by mixing the assay buffer with tracer and addition of the protein of interest at concentration equal to the measured $K_D$ value.

2. For dose-response measurements, competing compounds can either be added by hand or by using small volume liquid handlers. If no liquid handlers are available, doses need to be added in amounts which can be precisely pipetted by hand (≥ 1 µl) and therefore require compound addition prior to the transfer to microplates. Small volume liquid handler (e.g. Echo 550, Labcyte) are able to add nanoliters (nL) of compound solutions and therefore allow compound addition after transfer of the assay mix to the microplate. In order to obtain sufficient data points for reliable calculation of dose-dependent titration curves, measurement of 11 different concentrations (+ DMSO as a negative control) is recommended. These concentrations should be equally distributed in log10 units in the range of approx. 100 µM to 10 nM (highest concentration can be lowered for high affinity binders). For a 384-well plate transfer 10 µL per well and 5 µL if 1536-well plates are used.

3. After either mixing the compounds with the master mix and transferring them onto a micro titer plate (pipetting by hand) or adding the compounds to the master mix distributed in a high density microplate (automated dispensing), incubate the plate 20 min at RT.

4. Read the plate in a microplate reader equipped with a polarization filter (e.g. PHERAstar FSX equipped with the FP 590-50 675-50 675-50 filter module, BMG Labtech)

5. Plot the obtained mP values of the microplate reader against the log10(compound concentration). If a positive control is used with known high affinity, 0 % (DMSO negative control) and 100 % displacement (positive control) can be used for data normalization.

6. Use a three parameter fit (Hill Slope = -1.0) to fit the data, yielding an IC50 value for each dose-dependent titration.

7. To calculate the inhibitory constant ($K_i$) for each compound, the following equation published by Nikolovska-Coleska et al. [18] is used:

$$K_i = \frac{[I]_{50}}{\frac{[L]_{50}}{K_D} + \frac{[P]_0}{K_D} + 1} \tag{5}$$

Here, $[I]_{50}$ refers to the concentration of free inhibitor at 50 % inhibition, $[L]_{50}$ is the concentration of free tracer at 50 % inhibition, $[P]_0$ the concentration of free protein at 0 % and $K_D$ is the previously determined $K_D$ of the protein:tracer interaction.

For the calculation, the first parameter to be calculated is $[P]_0$ which requires the determined tracer $K_D$, the total concentration of tracer ($[L]_T$) and the total concentration of the protein used ($[P]_T$):

$$[P]_0 = \frac{-(K_D + [L]_T - [P]_T) - \sqrt{(K_D + [L]_T + [P]_T)^2 + 4 * [P]_T * K_D}}{2} \quad (6)$$

The obtained value for the concentration of free protein at 0 % inhibition can now be used for the calculation of the concentration of the free tracer at 0 % inhibition ($[L]_0$):

$$[L]_0 = \frac{[L]_T}{1 + \frac{[P]_0}{K_D}} \quad (7)$$

Next, the concentration of the protein:ligand complex at 0 % inhibition ($[PL]_0$) needs to be calculated using the following equation:

$$[PL]_0 = \frac{[P]_T}{\frac{K_D}{[L]_0} + 1} \quad (8)$$

Dividing the obtained concentration by 2, the concentration of the protein-ligand complex at 50 % inhibition ($[PL]_{50}$) can be calculated:

$$[PL]_{50} = \frac{[PL]_0}{2} \quad (9)$$

The value for $[PL]_{50}$ can now be used for the calculation of the concentration of free tracer at 50 % inhibition ($[L]_{50}$)

$$[L]_{50} = [L]_T - [PL]_{50} \tag{10}$$

Finally, the $IC_{50}$ obtained from the dose-dependent titration of the inhibitor in form of the turning point of the sigmoidal fit needs to be converted into the concentration of free inhibitor at 50 % inhibition ($[I]_{50}$) by using the following equation:

$$[I]_{50} = IC_{50} - [P]_T + K_D * \frac{[PL]_{50}}{[L]_{50}} + [PL]_{50} \tag{11}$$

Calculation of $[I]_{50}$ now allows for the calculation of the $K_i$ by using equation (5).

### 3.4. High throughput screening of compound libraries using fluorescence polarization

After measuring of control compounds or control peptides such as unlabeled p62-LIR in a dose-dependent manner, single concentration (one-shot) measurements can be used for high-throughput hit identification campaigns. The obtained mP value can also be used to estimate an affinity by comparison with the mP obtained from dose-dependent measurements. For high throughput screening, 1536-well plates are ideally used to obtain up to 100,000 data points per day. If measurement in 1536-well plate is not feasible, 384-well plates may be used reducing throughput.

1. Prepare an assay master mix by mixing the assay buffer with tracer and addition of the protein of interest at concentration equal to the determined $K_D$ value.
2. Prepare compounds by diluting compound stocks to equal concentrations (10 mM recommended). Determine the compound concentration for screening according to the expected hit affinity. For low affinity binders ($K_D$ range 10-1000 µM), screening at 100 µM is recommended, while screening at 10 µM neglects low affinity binders, mostly focusing on low µM to nM binders.

3. If no liquid handling system is available, prepare a master plate (e.g. 96-well plate) containing 1 µl compound and add master mix to obtain the desired concentration (e.g. add 100 µl master mix to 1 µl of 10 mM compound for screening 100 µM of the compound in the assay) before transferring the mixture in triplicates to the microplate. Using atomization, the master mix can be distributed onto the microplate followed by addition of the compounds in triplicates (e.g. add 50 nl of a 10 mM compound stock to 5 µl of master mix in a 1536-well plate).
4. Incubate the plate(s) at RT for 20 min.
5. Read the plate in a microplate reader equipped with a polarization filter (e.g. PHERAstar FSX equipped with the FP 590-50 675-50 675-50 filter module, BMG Labtech)
6. Using automation and stacker attachments on the microplate reader, one 1536-well plate can be measured in 20 min using dual channel detection (as available in the PHERAstar FSX platereader, BMG Labtech), allowing for 4600 data points in 1h and 110,000 data points in 24h.

## 4 Notes

1. No clones after Transformation: If no clones can be identified after transformation, either the DNA has insufficient purity/concentration or the cells were not handled correctly. To make sure that the plasmid is present in sufficient quality and amount, measure the concentration of the purified PCR product with a NanoDrop spectrophotometer (Thermo Fisher, or comparable). A purity check can be done by measurement of the A260/A280 ratio (expected between approx. 1.8 and 2) and the A260/A230 ratio (expected to be approx. 2.0 to 2.2). If the values are off or the concentration insufficient (e.g. < 20 ng/µl), re-purify the plasmid DNA. If the DNA is in the expected range, thaw another aliquot of cells. For transformation, it is crucial to handle the cells carefully and always on ice. Do not vortex the cells and avoid large shear forces when pipetting. Make sure, the recovery incubation directly after transformation for 1 h is carried out without antibiotics and the plates contain the correct antibiotic in the correct concentration.

2. Insufficient cell lysis: incomplete lysis of the cells can lead to a drastic loss in yield. Make sure that the cell lysate does not contain any visual particles or cell clumps and should be coloured evenly throughout the whole beaker. Additionally, high viscosity indicates genomic DNA in the lysate and can be resolved through addition of more DNase I.

3. No protein after affinity chromatography: It can happen that no protein elutes after affinity chromatography. Since all human LC3/GABARAPs were already validated to be successfully expressed and purified in the described assay setup, no construct errors can be expected. To ensure high amounts of soluble protein, keep the lysate constantly on ice and isolate the protein within one day. Do not store the lysate and ensure the correct pH value of the used buffers. Ensure, that the wash buffer does not contain too high imidazole concentrations which lead to premature elution. Lastly, check for protein expression by comparison of all fractions on an SDS-PAGE gel. If no band is found, the protein expression might not be induced by using IPTG.

4. Precipitated protein after dialysis: Some of the proteins tend precipitate during dialysis. One reason is the instability of the constructs induced by the $His_6$-tag. The stability is therefore drastically increased after cleavage but not during cleavage. Additionally, the TEV protease can also precipitate during this procedure. Due to the high yield of the human LC3/GABARAPs expression constructs, a limited extend of precipitation is of no concern. In this case, centrifugation of the suspension, followed by filtration through 0.22 µm filters and subsequent size exclusion chromatography will recover all functional protein from the suspension and make them suitable for the biophysical assays.

5. No binding of the tracer observed: Since the interaction of the described p62-LIR-based peptide was already successfully shown [14], missing binding detection is most likely driven by the assay setup or the protein quality. Ensure protein identity through SDS-PAGE and/or mass spectrometry. Measure protein concentration in triplicates using NanoDrop spectrophotometer (Thermo Fisher, or comparable) or carry out a BCA assay [19]. Ensure, that the correct filters are installed in the microplate reader and include positive (peptide) and negative (DMSO) controls. Additionally, ensure that Tween-20 was added to the assay buffer.

6. No displacement of the tracer: Generally, no displacement of the tracer indicates that the compound used for displacement is not competing with the tracer, meaning undetectable binding at the identical binding site. However, fluorescent compounds interfere with the tracer signal, leading to wrong readouts. Furthermore, insoluble compounds may also impair the readout of the assay and/or lead to protein precipitation.

7. Off-scale values during high throughput screening: It was already observed that the one-shot experiments lead to values which are completely off-scale or show 100 % displacement without interaction. In that cause, check the compound colour which mostly occur to be colourful. Since the compound colour cannot be influenced, interfering compounds must be excluded or screened through a technique which is not based on optical detection.

Acknowledgement

MPS, JD, SK and VVR are grateful for support by the Structural Genomics Consortium (SGC), a registered charity (no: 1097737) that receives funds from Bayer AG, Boehringer Ingelheim, Bristol Myers Squibb, Genentech, Genome Canada through Ontario Genomics Institute, EU/EFPIA/OICR/McGill/KTH/Diamond Innovative Medicines Initiative 2 Joint Undertaking [EUbOPEN grant 875510], Janssen, Merck KGaA, Pfizer and Takeda and by the German Cancer Research Center DKTK and the Frankfurt Cancer Institute (FCI). MPS is funded by the Deutsche Forschungsgemeinschaft (DFG, German Research Foundation), CRC1430 (Project-ID 424228829).

**Figure 1**

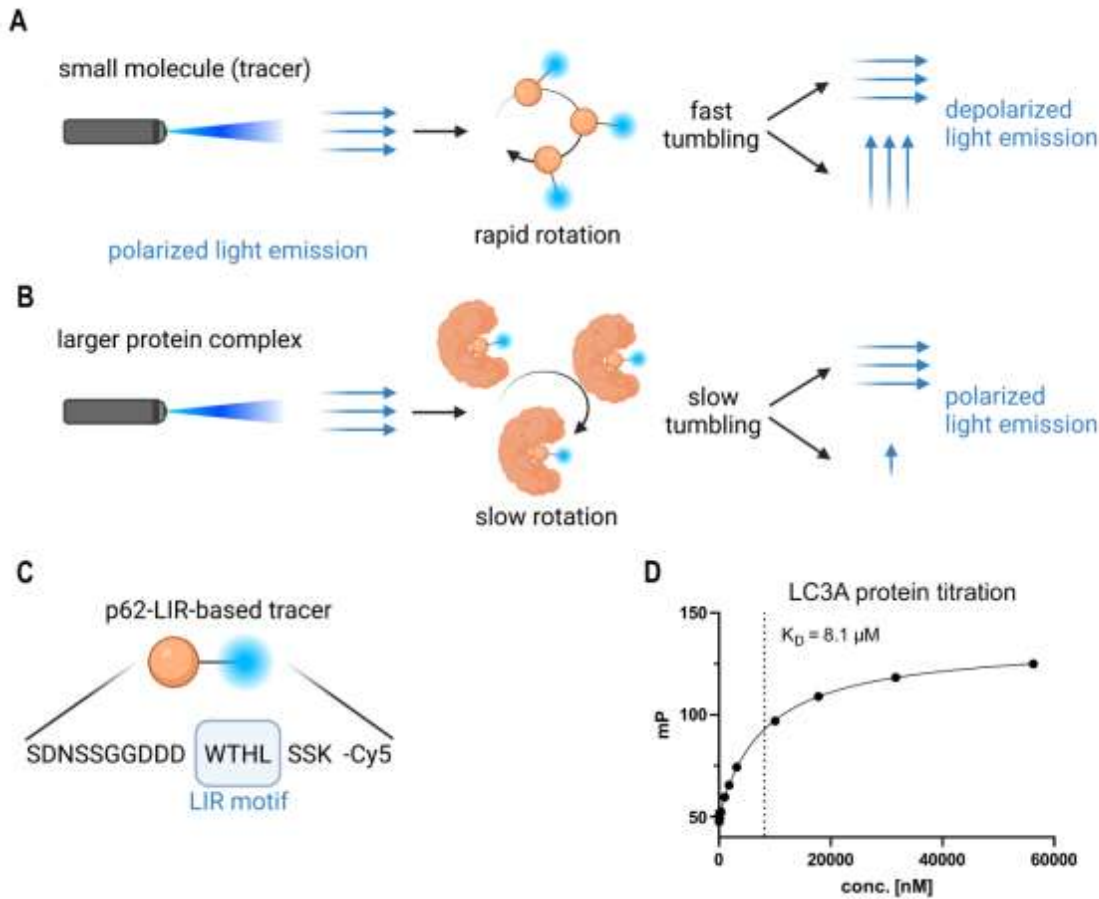

**Figure 1: Principle of the fluorescence polarization assay.** (**a**) and (**b**) the different mechanisms of free (**a**) and protein-bound (**b**) tracer. (**a**) The unbound tracer is excited by polarized light and fast tumbling leads to the emission of depolarized light. (**b**) binding to a protein partner increases the overall mass of the fluorophore-protein complex leading to slow tumbling and therefore lower emission of depolarized light. (**c**) Schematic overview of the p62-LIR-derived tracer, used for fluorescence polarization assays for LC3/GABARAPs. D) Exemplary protein titration curve obtained from LC3A titration against the p62-LIR tracer with the $K_D$ value determined according to 3.2 is given on the top.

**Figure 2**

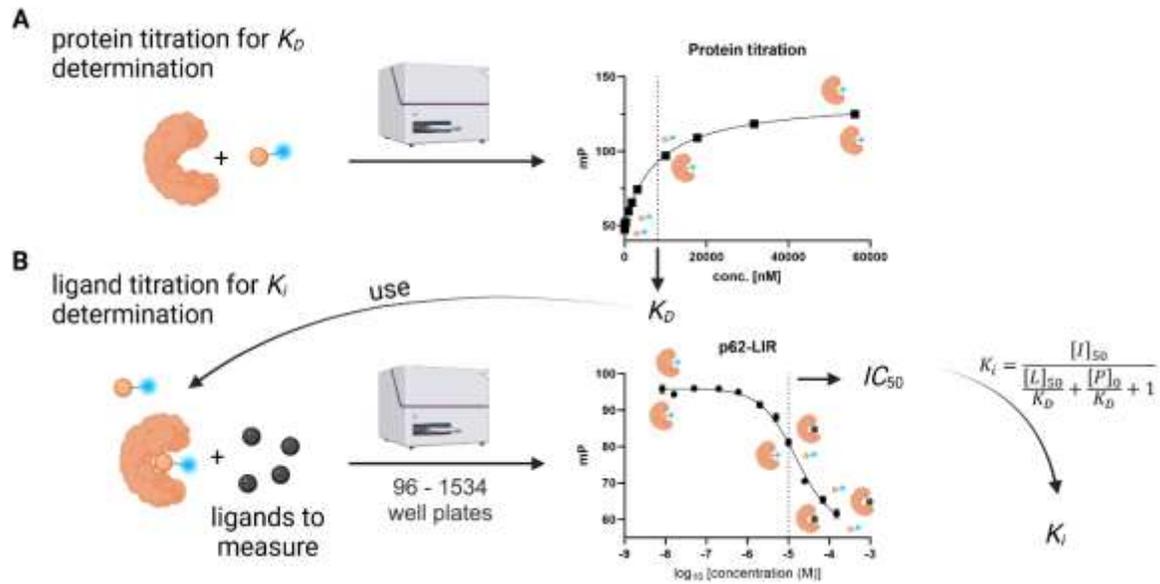

**Figure 2: Scheme of the high throughput fluorescence polarization assay.** (**a**) Protein titration to determine $K_D$ and optimal tracer concentration. (**b**) Determination of $K_i$ values in high throughput mode.